# Parameterization of Theoretical Methods in the Calculation of Nano-expulsion Energies

Aned de Leon[*]

*Instituto de Ciencias Nucleares, Universidad Nacional Autónoma de México, México D.F.*

*Abstract:*

The present work encompasses an investigation of the adsorption properties of small dimer species onto a zigzag single walled carbón nanotube (ZNT for short). To perform this analysis we have considered the zigzag (8, 0) system with a length of 13.5 Å that has a diameter of 6.3 Å. Various theoretical DFT and *ab initio* approaches were used to attempt to parameterize an ideal method for the calculation of expulsion energies for nanotube species.

*Keywords*: Zigzag SWNT; Dimer species, Adsorption; DFT; *Ab Initio*.

---

[*]To whom correspondence should be addressed.
E-mail: aned.leon@nucleares.unam.mx (A. de Leon)
Tel: +52-55-5646-0162

**1. Introduction**

The chemical and physical properties of single walled nanotubes (SWNT) has been the subject of recent studies[1]. Several investigations researchers have depicted that the binding energy of around 12 kcal/mol for the adsorption of hydrogen to the walls of nanotubes which reveal an affinity of the molecules for these carbon clusters. Interesting works also denote the threshold limits on the size of AlN SWNT systems[2] as well as the shortest feasible sizes of theses tubes[3].

Carbon nanotubes reserve many potentially important functions such as gas sensors, molecular field transistors and other materials science applications. The major challenges associated with these systems are their synthesis which is cumbersome. The products obtained from such studies demonstrate that the integration of SWNT species and fullerenes can also lead to pea-pod structures with aspiring applications[4].

Our group has successfully carried out theoretical calculations suggesting that fullerenes can interact with biochemical species (such as amino acids) with strong non-bonding interactions[5]. Later we applied these concepts to the differentiation of encapsulation process of dimer molecules to the surface and the interior[6]. This latter methodology was applied to the molecular engineering of a novel "scorpion" shaped SWNT-graphene sheet molecular trap[7]. SWNT species have been shown to encapsulate small polar species on their surfaces[8]. However, encapsulation of molecules leads to unusual modifications of the surface electron density of the nanostructure being studied[9].

It is believed that dispersion forces are the primary force in the interaction between the fullerene and the polar molecules upon encapsulation. The forces that sustain interactions between internally placed molecular dimers and the SWNT species are London forces.

These so-called effects are long-range attractions that act between separated molecules in the absence of charges or electric moments. The forces can be accounted for quantum mechanically, but primarily arise from the interplay between electrons belonging to the densities of two otherwise non-interacting molecules or atoms.

These dispersion interactions are important and other investigations reveal that the physical and electronic properties of the adsorbed materials and nanotubes change upon such proximity between separated entities[10-13]. These changes upon polar molecular encapsulation[5] can partially be explained by Stone-Wales defects that occur on 5-7 ring defects[14]. The present manuscript is oriented at characterizing the physical properties of the SWNT materials using a variety of DFT and *ab initio* approaches in order to make suitable recommendations for future reference.

## 2. Computational Methods

Previous theoretical calculations performed on the species studied in this work were performed with the BLYP/DND theoretical approach with a double numerical polarization and diffuse basis sets implemented in the DMol$^3$ numerical-based density functional computer software[15]. The present study implements the use of the GAUSSIAN03 suite of codes[16] to carry out a series of calculations. Four specific combinations were performed using the following method combinations:

   a. MP2/6-31++G**:UFF
   b. MP2/6-31++G**:AM1
   c. MP2/6-31++G**:UHF/3-21G*
   d. MPW1B95/6-31++G**:MPW1B95/3-21G*

In these notations MP2 is the second-order Møller-Plesset perturbation theory[17], AM1 is the Austin Model 1 semi-empirical method[18], UFF is a molecular mechanics approach[19], UHF is the Hartree-Fock method[20], and MPW1B95 is a meta-hybrid density functional theory (DFT) method[21].

The ONIOM approach was implemented[22] that incorporates a high level of theory for a section of the molecule and a lower level method for the remainder of the complex studied. The present case considered the high level of theory for the encapsulated dimer species and the low level method for the SWNT. Scheme 1 graphically depicts the distinction between the levels used in the calculations. In this work the $(H_2O)_2$, $(H_2S)_2$, $(H_2O)_2$, $(HCl)_2$, $(HF)_2$, $(NH_3)_2$ dimer systems were used in the studies.

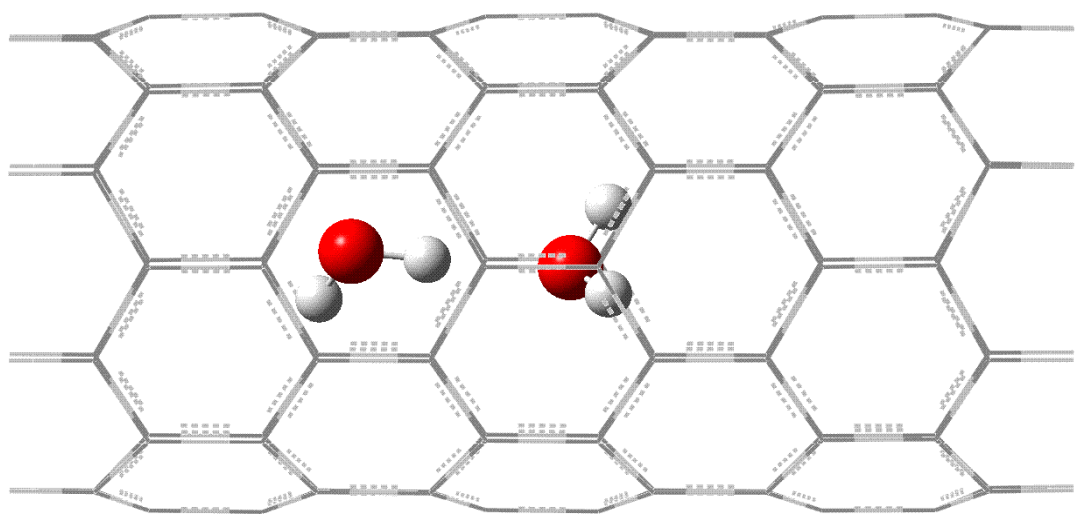

**Scheme 1.** ONIOM schematic diagram where the wireframe denotes the low-level SWNT calculation and the ball and stick model is the high-level dimer computation.

**3. Results and Discussion**

Table 1 depicts the dissociation energies (in kcal/mol) of the dimer species inside the zigzag (8, 0) species and outside (the values in parenthesis). The ONIOM approach was used in these computations whereby the following high level/low level combinations were applied: a). MP2/6-31++G**:UFF, b). MP2/6-31++G**:AM1, c).MP2/6-31++G**:UHF/3-21G*. d). MPW1B95/6-31++G**:MPW1B95/3-21G*. Table 2 shows the differences in energies between adsorption of the dimer species inside versus outside of the SWNT also in kcal/mol (using the same ONIOM scheme as in Table 1).

Figure 1 shows geometrical parameters for the endohedral polar molecular encapsulations is shown whereby bond lengths are in angstroms (Å) and bond lengths in degrees (º) where the values shown are the MP2/6-31++G**:UHF/3-21G* parameters and those in parenthesis are the MPW1B95/6-31++G**:MPW1B95/3-21G* results. The geometrical parameters for the externally adsorbed species have been presented in other works[5] and will not be repeated here.

*3.1 $(H_2O)_2$ Dimer Species*

For the first case we will discuss the water dimer encapsulated case has similar geometrical parameters are quite similar in both cases, with the intermolecular distances of 2.48 and 2.37 Å, respectively between the MP2 method and the DFT method are observed. At the first method using the MP2/6-31++G**:UFF method the $(H_2O)_2$ dimer is stabilized by around -6.03 kcal/mol on the interior of the SWNT and -1.71 kcal/mol on the exterior. For the next method (namely MP2/6-31++G**:AM1) the semi-empirical correction to UFF for the SWNT leads to values of 17.18 and 0.60 kcal/mol for adsorption on the interior and externally to the SWNT surface.

Improvement of the AM1 method to the MP2/6-31++G**:UHF/3-21G* that now includes HF potentials for the SWNT causes the interior/exterior adsorption energies to -2.96 and 11.5 kcal/mol, respectively. Finally, using the MPW1B95/6-31++G**:MPW1B95/3-21G* DFT approach we obtain values of -27.62 and -8.74 kcal/mol for adsorption to the interior and exterior, respectively. Overall, we can observe from Table 2 that the most stable configurations by these calculations is to the interior of the SWNT. With the MP2/UFF approach leads to a separation of -33.2 kcal/mol and when the HF method is applied for the SWNT calculation a value of -41.46 kcal/mol is computed. In the DFT method the stability is around -18 kcal/mol (with respect to the exterior adsorption process).

*3.2 $(H_2S)_2$ Dimer Species*

The geometrical parameters for this complex are generally well defined with the DFT and MP2 approaches used here. The primary deviations lie in the distance to the walls of the SWNT and the intermolecular hydrogen bond. These two quantities are difficult to model and can lead to computational error when comparing DFT to an *ab initio* approach.

Table 1 shows that in the cases considered the dissociation energies for the complexes at the MP2/UFF level of theory are indeed quite endothermic for the encapsulation process which is consistent among all methods used within the MP2 framework. At the MP2/HF method the interior and exterior adsorption energies are 35.6 and -14.8 kcal/mol, respectively. By use of the DFT method the interior and exterior adsorption energies are - 28.6 and -1.86 kcal/mol, respectively. Table 2 shows that the most marked difference is in the MP2/UFF method as well as the DFT approaches used. One must be conscious of the fact that the UFF method does not take into account any type of electron correlation effects and can lead to drastic errors in the calculation of the SWNT system. Additionally, further correlation is important for a full electronic diagram of the observed chemical behavior to be properly discussed.

*3.3 $(HCl)_2$ Dimer Species*

From the figure we can observe that the geometries in this case between the MP2 and DFT schemes are well correlated with the largest separation being between the SWNT wall and the internally encapsulated molecule. Again the MP2 methods yield positive encapsulation energies with favorable external adsorption quantities. At all levels of theory the MP2 adsorption energies are lower than the DFT values which suggest favorable encapsulation energies. We can see from the table that the largest differences are observed again for the MP2/UFF method which can be accounted for due to errors in the correlation energies.

*3.4 $(HF)_2$ Dimer Species*

Similarly as in the previous example the geometries obtained in the MP2 and DFT approaches are consistent with the intermolecular separations differing marginally. In all calculation schemes in this example the encapsulation energies are favorable as can be seen from the table. With further correlation as can be seen in the MP2/HF and DFT ONIOM method the energies of adsorption are exothermic in both the interiorly and exteriorly adsorbed species. It is apparent from the calculations that adsorption to the interior of more favorable than on the exterior.

*3.5 $(NH_3)_2$ Dimer Species*

The stationary geometries of these species as calculated by MP2 and the DFT approaches are in agreement except in the description of the $NH_3$---$NH_3$ bond length. This of course can be attributed to differences in the dispersion interactions with vary in the two schemes. The MP2 schemes provide positive dissociation energies for the adsorption energies with the process leading to a favorable mechanism in the DFT calculations. Table 2 shows that the deviations in the MP2/UFF method are most dramatic and the MP2 and DFT relative differences being similar. The primary conclusion that can be drawn is that by using a correlated DFT method the adsorption energies on the surface are not as favorable as those on the interior.

**4. Conclusions**

Throughout this study we have observed that the adsorption properties of the small polar molecules to the surface of the SWNT and the interior can be modeled using a variety of approaches. Namely, the MP2 *ab initio* method which is excellent at predicting the physical properties of molecular structures can be combined with fast molecular mechanics (UFF) methods, semi-empirical (AM1) method, and the hartree-fock (HF) method. While, further improvements can be extended to a full MP2 optimization scheme of the SWNT this is a tedious task that is computationally expensive. Additionally, relatively fast DFT computations using the MPW1B95/6-31++G**:MPW1B95/3-21G* method have been shown to reduce the basis set superposition error (BSSE) as well as account for dispersion[21].

From a physical point of view the MP2/HF combined method reveals that the most stable conformation is the $(HF)_2$ species followed by $(NH_3)_2$ and with the DFT method the most stable dimers are the $(H_2O)_2$ and $(H_2S)_2$ molecular complexes. The $(NH_3)_2$ species is also quite stable but the external adsorption energies are somewhat depleted.

When selecting a method for the calculations it is important to use ONIOM computational schemes that account for BSSE as well as dispersion. Therefore, while the MP2/HF combination is recommended, at times DFT schemes with diffuse basis sets should be used in order to yield results in reasonable computation times. However, one must be careful in analyzing the results since further testing is indeed necessary. While the DFT methods reveal that the adsorption to the interior is stable this can is different in the MP2 approaches. Possible, the deviation can be accounted for that a lack of correlation in the SWNT needs to be taken into account. It is our assertion that the calculations should be of use for calculations involving ONIOM and other hybrid quantum mechanical (QM) techniques.

**Acknowledgments**

We are grateful to CONACYT and the UNAM for valuable financial resources with particular appreciation to the DGSCA for supercomputer time.

## Table and Figure Captions

**Table 1.** Dissociation energies of the adsorption properties of the dimer species inside the zigzag (8, 0) species and outside (the values in parenthesis) in kcal/mol. In the calculations the ONIOM approach was used whereby the following high level/low level combinations were applied: a). MP2/6-31++G**:UFF, b). MP2/6-31++G**:AM1, c).MP2/6-31++G**:UHF/3-21G*. d). MPW1B95/6-31++G**:MPW1B95/3-21G*

**Table 2.** Deviation in energies between adsorption of the dimer species inside versus outside of the SWNT in kcal/mol. Again, in the computations the ONIOM approach was used whereby the following high level/low level combinations were applied: a). MP2/6-31++G**:UFF, b). MP2/6-31++G**:AM1, c).MP2/6-31++G**:UHF/3-21G*. d). MPW1B95/6-31++G**:MPW1B95/3-21G*

**Figure 1.** Selected geometrical parameters are depicted whereby bond lengths are in angstroms (Å) and bond lengths in degrees (º) where the values shown are the MP2/6-31++G**:UHF/3-21G* parameters and those in parenthesis are the MPW1B95/6-31++G**:MPW1B95/3-21G* results.

| Dimer Species | MP2/UFF[a] | MP2/AM1[b] | MP2/UHF[c] | MPW1B95[d] |
|---|---|---|---|---|
| (H$_2$O)$_2$ | -6.03 | 17.18 | -2.96 | -27.62 |
|  | (-1.71) | (0.60) | (11.50) | (-8.74) |
| (H$_2$S)$_2$ | 21.62 | 40.70 | 35.59 | -28.60 |
|  | (-5.56) | (-0.61) | (-14.78) | (-1.86) |
| (HCl)$_2$ | 1.80 | 200.90 | 21.84 | -27.90 |
|  | (-4.16) | (-0.46) | (-1.39) | (-4.29) |
| (HF)$_2$ | -10.33 | 22.33 | -6.72 | -17.45 |
|  | (15.93) | (3.83) | (-3.91) | (-1.18) |
| (NH$_3$)$_2$ | 1.18 | 14.92 | 7.95 | -24.56 |
|  | (-4.98) | (0.85) | (-0.91) | (-3.55) |

**Table 1.**

| Dimer Species | MP2/UFF[a] | MP2/AM1[b] | MP2/UHF[c] | MPW1B95[d] |
|---|---|---|---|---|
| $(H_2O)_2$ | -33.21 | -3.70 | -41.46 | -17.99 |
| $(H_2S)_2$ | -102.34 | 15.66 | 6.86 | -20.36 |
| $(HCl)_2$ | -103.28 | 176.31 | -22.11 | -18.78 |
| $(HF)_2$ | -51.28 | -4.56 | -31.24 | -17.28 |
| $(NH_3)_2$ | -134.72 | -7.43 | -31.46 | -20.07 |

**Table 2.**

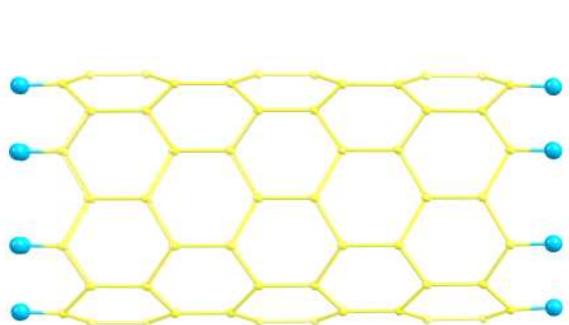

**SWNT**

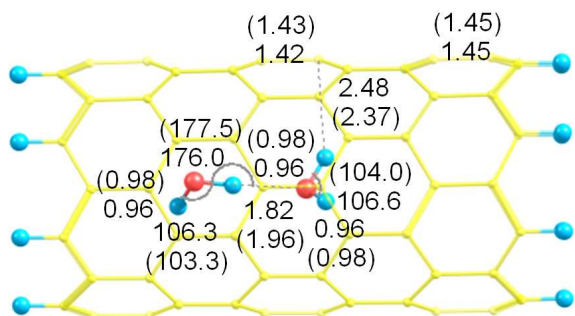

**(H₂O)₂**

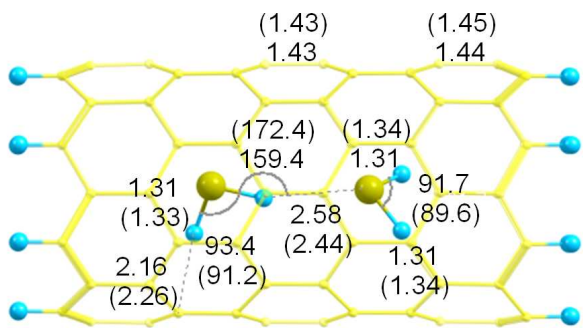

**(H₂S)₂**

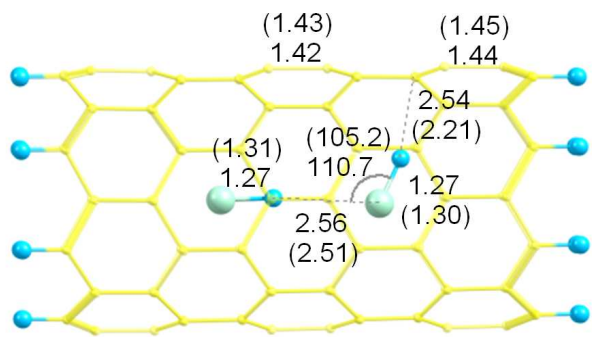

**(HCl)₂**

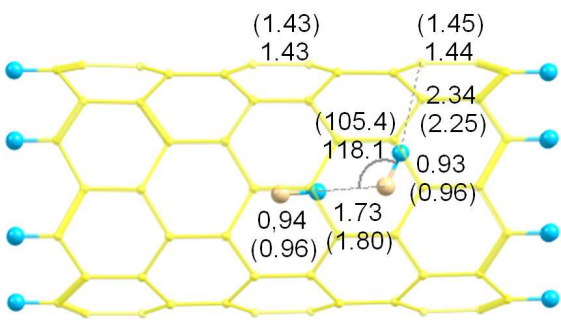

**(HF)₂**

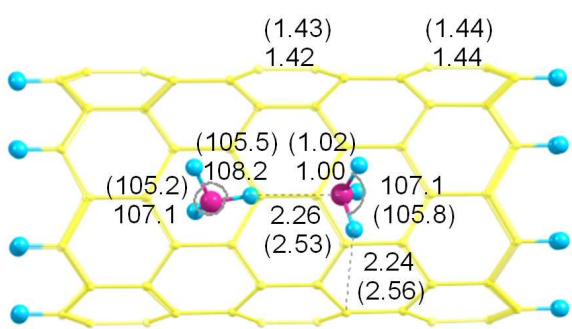

**(NH₃)₂**

**Figure 1.**